\documentclass[10pt,conference]{IEEEtran}
\IEEEoverridecommandlockouts
\usepackage{comment}
\usepackage{multirow}
\usepackage{booktabs}

\usepackage{cite}
\usepackage{amsmath,amssymb,amsfonts}
\usepackage{algorithmic}
\usepackage{graphicx}
\usepackage{textcomp}
\usepackage{xcolor}
\def\BibTeX{{\rm B\kern-.05em{\sc i\kern-.025em b}\kern-.08em
    T\kern-.1667em\lower.7ex\hbox{E}\kern-.125emX}}

\usepackage{marvosym}  

\begin{document}

\title{Audio Spatially-Guided Fusion for Audio-Visual Navigation}
\author{
Xinyu Zhou$^{1,2,3}$, and Yinfeng Yu$^{1,2,3}$$^{,\mbox{\Letter}}$%
\thanks{$^{\mbox{\Letter}}$Yinfeng Yu is the corresponding author(E-mail: yuyinfeng@xju.edu.cn).}%
\\
$^1$Joint Research Laboratory for Embodied Intelligence, Xinjiang University\\
$^2$Joint International Research Laboratory of Silk Road Multilingual Cognitive Computing, Xinjiang University\\
$^3$School of Computer Science and Technology, Xinjiang University, Urumqi 830017, China% 
}

\maketitle
\suppressfloats[t]

\begin{abstract}
Audio-visual Navigation refers to an agent utilizing visual and auditory information in complex 3D environments to accomplish target localization and path planning, thereby achieving autonomous navigation. The core challenge of this task lies in the following: how the agent can break free from the dependence on training data and achieve autonomous navigation with good generalization performance when facing changes in environments and sound sources. To address this challenge, we propose an Audio Spatially-Guided Fusion for Audio-Visual Navigation method. First, we design an audio spatial feature encoder, which adaptively extracts target-related spatial state information through an audio intensity attention mechanism; based on this, we introduce an Audio Spatial State Guided Fusion (ASGF) to achieve dynamic alignment and adaptive fusion of multimodal features, effectively alleviating noise interference caused by perceptual uncertainty. Experimental results on the Replica and Matterport3D datasets indicate that our method is particularly effective on unheard tasks, demonstrating improved generalization under unknown sound source distributions.
\end{abstract}


\section{Introduction}

In recent years, the rapid development of emerging industries such as autonomous driving~\cite{b1} and smart home systems~\cite{b2} has motivated researchers to explore new navigation paradigms, among which embodied navigation~\cite{b3,b4,b5} has attracted increasing attention. Audio-Visual Navigation (AVN)~\cite{b6,b7} is a significant research direction in the field of embodied navigation. It aims to enable agents to localize targets and navigate autonomously using audio-visual observations in complex environments, with broad application potential in scenarios such as rescue and household service.
Current audio-visual navigation algorithms can achieve a navigation success rate of 98\% in tests with heard sounds. However, their performance drops significantly to 52\% in tests with unheard sounds. This indicates that agents in current research have obvious deficiencies in generalization ability across sound sources and scenarios. Yet real application scenarios involve diverse types of sounds. We cannot achieve full coverage with training data. Therefore, improving the generalization ability of navigation models is a crucial key issue.
\begin{figure}[!htbp]
    \centering
    \includegraphics[width=\columnwidth]{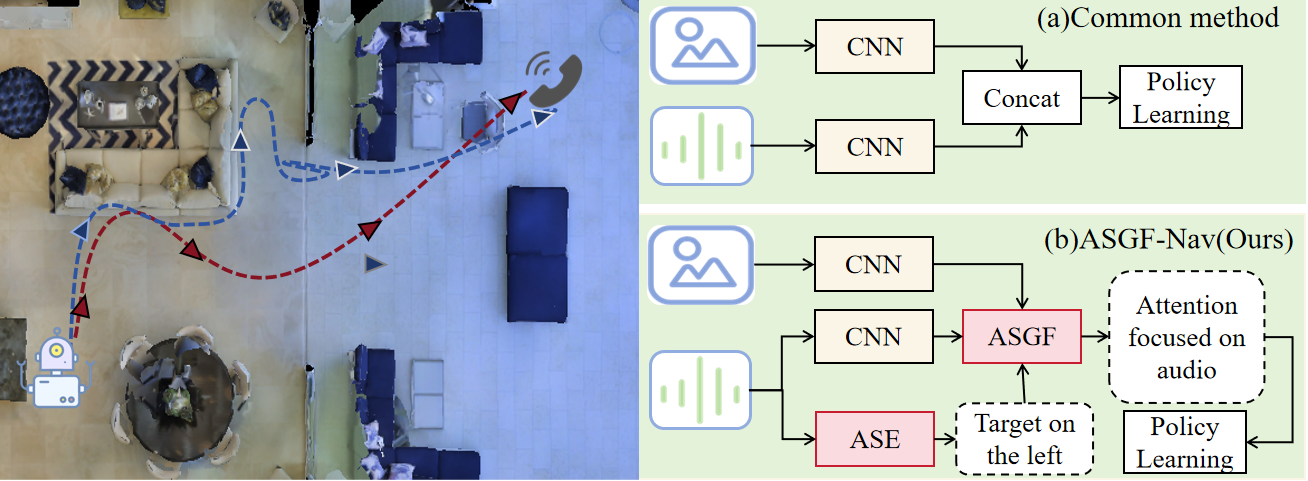}
    \caption{Comparison of navigation trajectories and model architectures. Left: Top-down view of navigation trajectories in an indoor scene (blue: our ASGF-Nav; red: baseline method). Right: (a) Common audio-visual navigation method with simple feature concatenation; (b) Our ASGF-Nav model, which integrates the ASE module to extract spatial information from audio and uses the ASGF module to dynamically fuse audio-visual features.}
    \label{fig:intro}
\end{figure}

%整体图
\begin{figure*}[!t]
    \centering
    \includegraphics[width=\textwidth]{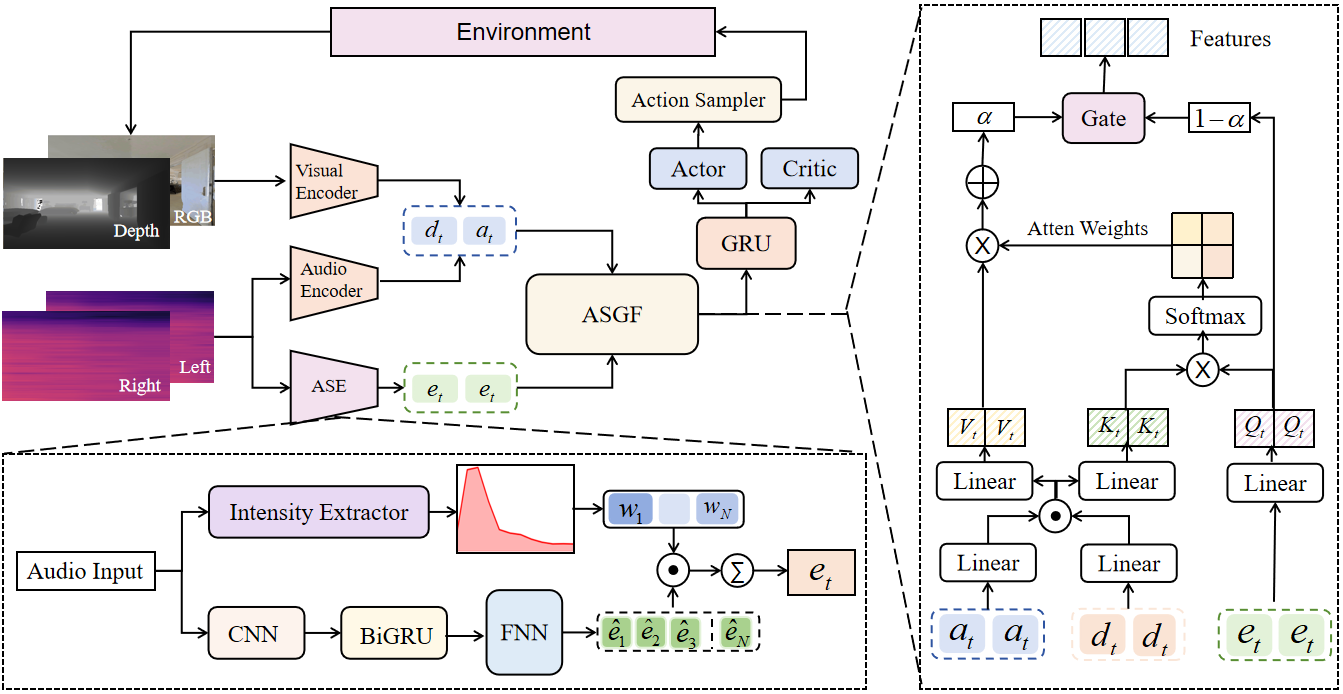}
    \caption{Model architecture. Our audio spatially-guided fusion for audio-visual navigation model (ASGF-Nav) uses the ASE module to extract implicit spatial state information from binaural spectrograms, and guides the dynamic fusion of visual and audio features in the ASGF module to provide a basis for policy selection.}
    \label{fig:model}
\end{figure*}

Through the analysis of existing research results, we can attribute the limited generalization performance to the following two points:
1) Insufficient audio feature extraction capability, where simple convolutional encoders struggle to model temporal dependencies and fully exploit spatial cues in audio~\cite{b24,b28,b31}. 2) limited flexibility in multimodal fusion, where fixed-weight fusion cannot adaptively adjust the contributions of audio and visual information under changing scenes or sound categories~\cite{b25,b26}.

In 3D environments, sound propagation is affected by factors such as relative direction, propagation distance, and environmental structure. Such space-related information is implicitly embedded in audio signals. Based on this understanding, we propose a cross-modal fusion framework guided by audio spatial states. The agent learns spatial state information from audio, and uses this information to guide the dynamic fusion of visual and audio features. In this way, the agent is no longer limited to memorizing simple correspondences between sounds and environments. Instead, it can exploit the spatial constraints provided by audio, achieving fast localization and efficient autonomous navigation of target sound sources. Experimental results show that compared with existing research, our proposed model achieves clear improvements on Unheard tasks and exhibits stronger generalization ability.

In summary, our main contributions are as follows:
\begin{itemize}
\item We propose a temporal-aware audio feature extraction module based on CRNN, combined with audio intensity attention, to effectively extract spatial state information from audio.
\item We propose a cross-modal fusion framework guided by audio spatial states, which adaptively coordinates the weights of audio and visual modalities in different scenarios.
\item We conduct experiments in complex 3D environments and compare with multiple baseline models, proving that the proposed method has good generalization performance.
\end{itemize}.

\section{Related Work}
\subsection{Audio-Visual Navigation}
Audio-Visual Navigation  requires agents to use visual and auditory observation information in complex 3D environments to autonomously plan paths and reach sound source targets~\cite{b22}. In recent years, existing studies have explored the audio-visual navigation task from different processing methods of application scenarios and observation information. SoundSpaces~\cite{b6} builds the first simulation platform for audio-visual navigation tasks based on the Replica~\cite{b7} and Matterport3D~\cite{b8} real indoor scene datasets, providing a unified experimental environment for audio-visual navigation methods. LLA~\cite{b9} adopts a phased perception and planning structure to enhance the stability during the policy learning process. AV-WAN~\cite{b10} models observation information as sound and visual maps, achieving efficient navigation by decomposing long-term goals into short-term goals. CAHM~\cite{b11} combines local geometric maps with historical observations to achieve tracking of slowly moving sound sources. SAVi~\cite{b12} further introduces semantic information to explicitly construct the correspondence between sound events and objects in the environment. SAAVN~\cite{b13} constructs a complex acoustic environment with interference and adversarial attacks to improve the robustness of the navigation agent. ORAN~\cite{b14} changes the traditional single-view observation paradigm by collecting information in an omnidirectional manner and employing knowledge distillation to achieve effective transfer of target navigation policies. Inspired by vision-language navigation, AVLEN~\cite{b15} and CAVEN~\cite{b16} introduce an Oracle to provide natural language instructions to the agent to assist it in locating navigation targets. More recently, AVN research has further explored collaborative acoustic perception, echo-enhanced navigation, multi-agent cooperation, and stereo-aware dynamic fusion strategies~\cite{b23,b24,b27,b28}.Related embodied navigation studies have also investigated enhanced object-level perception to improve policy learning~\cite{b30}. Despite these advances, how to better exploit audio information for robust navigation in complex and unseen environments remains an open problem.

\subsection{Feature Fusion}
In audio-visual navigation tasks, effective navigation depends on jointly leveraging visual and auditory modalities. From the perspective of fusion strategies, most existing approaches adopt early-stage feature-level fusion with fixed weights. Specifically, methods such as SoundSpaces~\cite{b6}, LLA~\cite{b9}, and AV-WAN~\cite{b10} independently encode visual and audio features and then simply concatenate or linearly combine them as input to the policy network. However, such fusion strategies are unable to dynamically adjust the contribution of each modality in response to environmental changes or perceptual uncertainty. With the advancement of attention mechanisms, an increasing number of works have introduced self-attention, cross-modal attention, and dynamic fusion into the fusion process~\cite{b17,b28,b25,b26}. For example, FSAAVN~\cite{b17} employs a self-attention module to achieve context-aware fusion of audio and visual features. SAVi~\cite{b12} combines Transformer-based architectures with memory modules to address the issue of information loss under sparse or intermittent sound conditions. Related dynamic gating and end-to-end audio-visual fusion ideas have also been explored in broader audio-visual learning tasks~\cite{b29,b31}. Nevertheless, agent performance still degrades significantly in the~\textit{Unheard} environment.

\section{Method}
We design an audio spatial state-guided cross-modal fusion framework to enhance the generalization capability of the model. As illustrated in Fig.~\ref{fig:model}, RGB and depth images are used as visual inputs, while binaural spectrograms are used as auditory inputs. Visual and auditory observations are first processed by convolutional neural networks to extract modality-specific features. For the auditory modality, we further introduce an Audio Spatial State Encoder (ASE) to explicitly learn spatial state information embedded in audio signals. The extracted visual features, audio features, and audio spatial state representations are then fed into the Audio Spatial State-Guided Cross-Modal Fusion (ASGF) module for feature integration. The fused features are recursively combined with historical hidden states through a gated recurrent unit (GRU), producing state representations enriched with temporal contextual information. The fused features are finally fed into an Actor-Critic network for policy learning and value estimation.

%ASE图
% ASE module illustration

\subsection{Observations Encoder}\label{AA}
At each time step $t$, two independent CNNs are used as the visual and audio encoders to extract visual and auditory features, denoted by $D_t \in \mathbb{R}^{N_d}$ and $A_t \in \mathbb{R}^{N_a}$, respectively.

We introduce an \emph{Audio Spatial State Encoder} (ASE) to model auditory observations and capture the spatial state information conveyed by audio signals, as shown in Fig.~\ref{fig:model}. Raw audio signals are first transformed into binaural spectrograms using the Short-Time Fourier Transform (STFT)~\cite{b19}. The resulting spectrograms are fed into a CNN to extract hierarchical time--frequency features while preserving the temporal axis as a sequence. A BiGRU is then applied to model temporal dependencies and produce frame-wise audio spatial embeddings
$\tilde{\mathbf{e}}_{t,i} \in \mathbb{R}^{N_e}$ for the current decision step $t$, where $i = 1,\dots,N$ indexes the frames in the audio segment, $N$ is the number of frames, and $N_e$ is the embedding dimension ($N_e = 512$ in our implementation).

Meanwhile, we calculate the sound intensity of each frame in the original binaural spectrogram, normalize the sound intensity via softmax, and obtain the frame-level attention weights.Specifically, the intensity of frame $i$ at decision step $t$ is defined as
\begin{equation}
I_{t,i} = \operatorname{mean}_{f,c} \bigl| X_t(f,i,c) \bigr|,
\end{equation}
where $X_t(f,i,c)$ denotes the magnitude at frequency bin $f$, frame index $i$, and channel $c$.
These intensity scores are converted into temporal attention weights via a softmax along the time axis:
\begin{equation}
w_{t,i} = 
\frac{\exp\bigl(I_{t,i}\bigr)}
     {\sum_{j=1}^{N} \exp\bigl(I_{t,j}\bigr)},
\label{eq:ase-attn}
\end{equation}
such that $\sum_{i=1}^{N} w_{t,i} = 1$.

Finally, the frame-wise embeddings are aggregated using the intensity-based attention weights to form a global audio spatial state representation at the current time step:
\begin{equation}
\mathbf{e}_t = \sum_{i=1}^{N} w_{t,i}\,\tilde{\mathbf{e}}_{t,i}
\in \mathbb{R}^{N_e},
\label{eq:ase-agg}
\end{equation}
Based on the audio intensity attention mechanism, the model can focus more on frames with high confidence and obtain a more robust representation of audio spatial states.

\begin{table*}[t]
\centering
\caption{Navigation Performance Comparison with other methods.}
\label{tab:main}
\setlength{\tabcolsep}{5pt}
\renewcommand{\arraystretch}{1.15}
\resizebox{\textwidth}{!}{%
\begin{tabular}{l ccc ccc ccc ccc}
\toprule
\multirow{3}{*}{\textbf{Method}}
& \multicolumn{6}{c}{\textbf{Replica}}
& \multicolumn{6}{c}{\textbf{Matterport3D}} \\
\cmidrule(lr){2-7}\cmidrule(lr){8-13}

& \multicolumn{3}{c}{\textbf{Heard}}
& \multicolumn{3}{c}{\textbf{Unheard}}
& \multicolumn{3}{c}{\textbf{Heard}}
& \multicolumn{3}{c}{\textbf{Unheard}} \\
\cmidrule(lr){2-4}\cmidrule(lr){5-7}
\cmidrule(lr){8-10}\cmidrule(lr){11-13}

& \textbf{SPL}$\uparrow$ & \textbf{SR}$\uparrow$ & \textbf{SNA}$\uparrow$
& \textbf{SPL}$\uparrow$ & \textbf{SR}$\uparrow$ & \textbf{SNA}$\uparrow$
& \textbf{SPL}$\uparrow$ & \textbf{SR}$\uparrow$ & \textbf{SNA}$\uparrow$
& \textbf{SPL}$\uparrow$ & \textbf{SR}$\uparrow$ & \textbf{SNA}$\uparrow$ \\
\midrule

Random Agent \cite{b10}
& 4.9 & 18.5 & 1.8
& 4.9 & 18.5 & 1.8
& 2.1 & 9.1 & 0.8
& 2.1 & 9.1 & 0.8 \\

Direction Follower \cite{b10}
& 54.7 & 72.0 & 41.1
& 11.1 & 17.2 & 8.4
& 32.3 & 41.2 & 23.8
& 13.9 & 18.0 & 10.7 \\

Gan et al. \cite{b9}
& 57.6 & 83.1 & 47.9
& 7.5 & 15.7 & 5.7
& 22.8 & 37.9 & 17.1
& 5.0 & 10.2 & 3.6 \\

AV-WAN \cite{b10}
& 86.6 & 98.7 & 70.7
& 34.7 & 52.8 & 27.1
& 72.3 & 93.6 & 54.8
& 40.9 & 56.7 & 30.6 \\

AGSA \cite{b28}
& 80.1 & 95.6 & 52.0
& 36.6 & 48.3 & 22.4
& 57.8 & 72.1 & 34.3
& 26.2 & 36.5 & 13.1 \\

SoundSpaces \cite{b6}
& 74.4 & 91.4 & 48.1
& 34.7 & 50.9 & 16.7
& 54.3 & 67.7 & 31.3
& 25.9 & 40.5 & 12.8 \\

\textbf{ASGF-Nav (Ours)}
& 82.7 & 94.5 & 62.7
& \textbf{63.3} & \textbf{76.5} & \textbf{36.9}
& 59.1 & 87.6 & 39.5
& \textbf{52.2} & \textbf{66.4} & 29.9 \\

\bottomrule
\end{tabular}%
}
\end{table*}

\subsection{Feature Fusion}\label{BB}
We propose the Audio Spatial State Guided Cross-Modal Fusion module for adaptive multimodal fusion, as illustrated in Fig.~\ref{fig:model}.

At time step $t$, the audio spatial state $e_t$, visual feature $d_t$, and auditory feature $a_t$ are first normalized and projected into a unified embedding space. The projected visual and auditory features are then concatenated to form a cross-modal memory representation, from which Key and Value representations are generated, while the audio spatial state serves as the Query. Specifically, the cross-attention computation is formulated as follows:
\begin{equation}
X_t = \big[ W_d\,\mathrm{LN}(d_t),\; W_a\,\mathrm{LN}(a_t) \big],
\end{equation}
\begin{equation}
K_t = X_t W_K, \quad V_t = X_t W_V,
\end{equation}
\begin{equation}
q_t = W_q\,\mathrm{LN}(e_t),
\end{equation}
\begin{equation}
c_t = \mathrm{Attn}(q_t, K_t, V_t),
\end{equation}
where $W_d \in \mathbb{R}^{D \times N_d}$ and $W_a \in \mathbb{R}^{D \times N_a}$ project $d_t$ and $a_t$ into the same $D$-dimensional latent space; $W_q \in \mathbb{R}^{D \times N_e}$, $W_K \in \mathbb{R}^{D \times D}$, and $W_V \in \mathbb{R}^{D \times D}$ are learnable projection matrices; and $c_t$ denotes the context representation obtained by selectively aggregating visual and auditory information conditioned on the audio spatial state.

The attended context is combined with the original audio spatial state through a residual connection:
\begin{equation}
\hat{h}_t = c_t + e_t,
\end{equation}

To handle varying modality reliability, we further introduce an adaptive gating mechanism. The gating network takes $e_t$ as input and outputs a scalar coefficient through a multi-layer perceptron (MLP):
\begin{equation}
\alpha_t = \sigma\big(\mathrm{MLP}(\mathrm{LN}(e_t))\big), \quad \alpha_t \in (0,1),
\end{equation}
where $\sigma(\cdot)$ denotes the Sigmoid function. The final fused state is obtained as
\begin{equation}
h_t = \mathrm{LN}_{\mathrm{out}}\big(\alpha_t \hat{h}_t + (1 - \alpha_t)e_t\big).
\end{equation}

\section{Experiments}
\subsection{Experimental Setup}\label{sec:exp-setup}
We conduct experiments on the public datasets Replica~\cite{b7}, Matterport3D~\cite{b8}, and the SoundSpaces audio simulation dataset. 
The Replica dataset~\cite{b7} contains 18 accurately scanned 3D indoor scenes, including apartments and offices. 
The Matterport3D~\cite{b8} dataset consists of 85 real-world indoor scenes, including residential environments. 
The audio perceived by the agent is synthesized by applying binaural room impulse responses, corresponding to the sound source directions, to the original sound signals.
In each episode, a scene is randomly selected, along with random starting positions for the agent and the target sound source. 
During navigation, the agent can execute four discrete actions: \{\textit{turn right}, \textit{turn left},  \textit{stop}, \textit{move forward}\}.

We evaluate the model under two task settings: 
\begin{enumerate}
\item{Heard-Sound}, where the target sound source remains the same (telephone) during training, validation, and testing; 
\item{Unheard-Sound}, where different sounds are used during training and testing to evaluate the model's generalization ability to unseen sounds. 
In both task settings, all test scenes are unseen during training.
\end{enumerate}

\subsection{Evaluation Metrics}\label{sec:exp-metrics}
In the experiments, we evaluate the navigation performance of different models using three metrics: 
\begin{enumerate}
\item{Success weighted by Number of Actions (SNA)}~\cite{b21}: the number of actions in successful navigation episodes is weighted for evaluating action efficiency during navigation; 
\item{Success Rate (SR)}: the proportion of episodes in which the agent successfully navigates to the target during testing; 
\item{Success weighted by Path Length (SPL)}: which weights successful navigation episodes based on the deviation between the actual traveled path and the shortest path.
\end{enumerate}

\begin{table}[t]
\centering
\caption{Ablation study of ASGF-Nav components on the \textit{Unheard} task.}
\label{tab:ablation_unheard}
\small
\setlength{\tabcolsep}{6pt}
\renewcommand{\arraystretch}{1.2}

\begin{tabular}{l ccc ccc}
\toprule
\multirow{2}{*}{\textbf{Method}}
& \multicolumn{3}{c}{\textbf{Replica (Unheard)}}
& \multicolumn{3}{c}{\textbf{Matterport3D (Unheard)}} \\
\cmidrule(lr){2-4}\cmidrule(lr){5-7}

& \textbf{SPL}$\uparrow$ & \textbf{SR}$\uparrow$ & \textbf{SNA}$\uparrow$
& \textbf{SPL}$\uparrow$ & \textbf{SR}$\uparrow$ & \textbf{SNA}$\uparrow$ \\
\midrule

SoundSpaces
& 34.7 & 50.9 & 16.7
& 25.9 & 40.5 & 12.8 \\

w/o ASGF
& 38.8 & 53.6 & 21.4
& 31.6 & 40.0 & 20.7 \\

w/o ASE
& 59.5 & 71.9 & 34.8
& 41.4 & 59.9 & 22.7 \\

\textbf{ASGF-Nav}
& \textbf{63.3} & \textbf{76.5} & \textbf{36.9}
& \textbf{52.2} & \textbf{66.4} & \textbf{29.9} \\

\bottomrule
\end{tabular}
\end{table}

%对比图
\begin{figure*}[t]
    \centering
    \includegraphics[width=\textwidth]{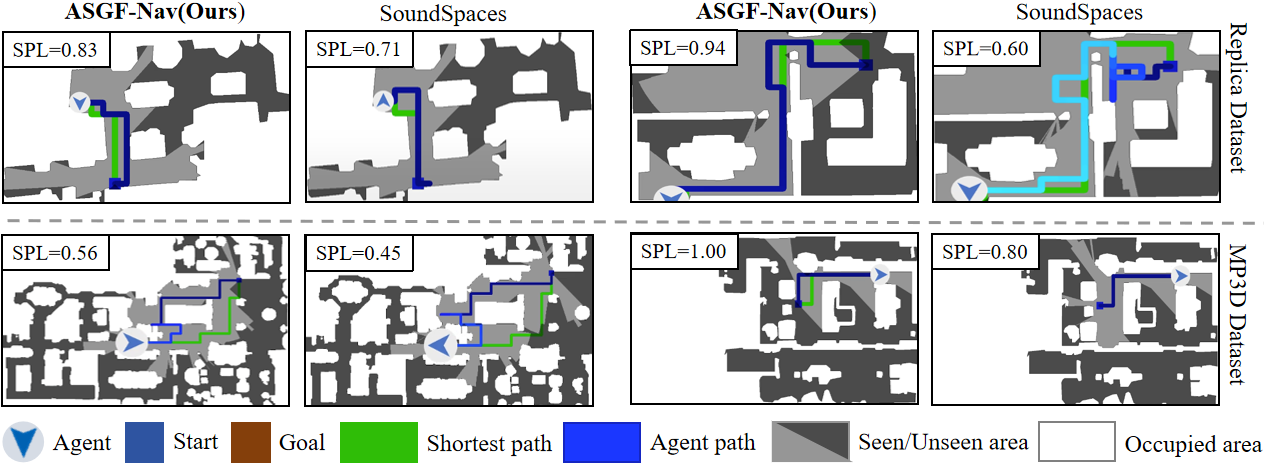}
    \caption{Top-down visualization of agent trajectories under the Unheard task. The color gradient from dark to light blue represents temporal progression. Compared with SoundSpaces, our method reaches the target with fewer detours.}
    \label{fig:trajectory}
\end{figure*}

%tsne图
\begin{figure}[t]
    \centering
    \includegraphics[width=\columnwidth]{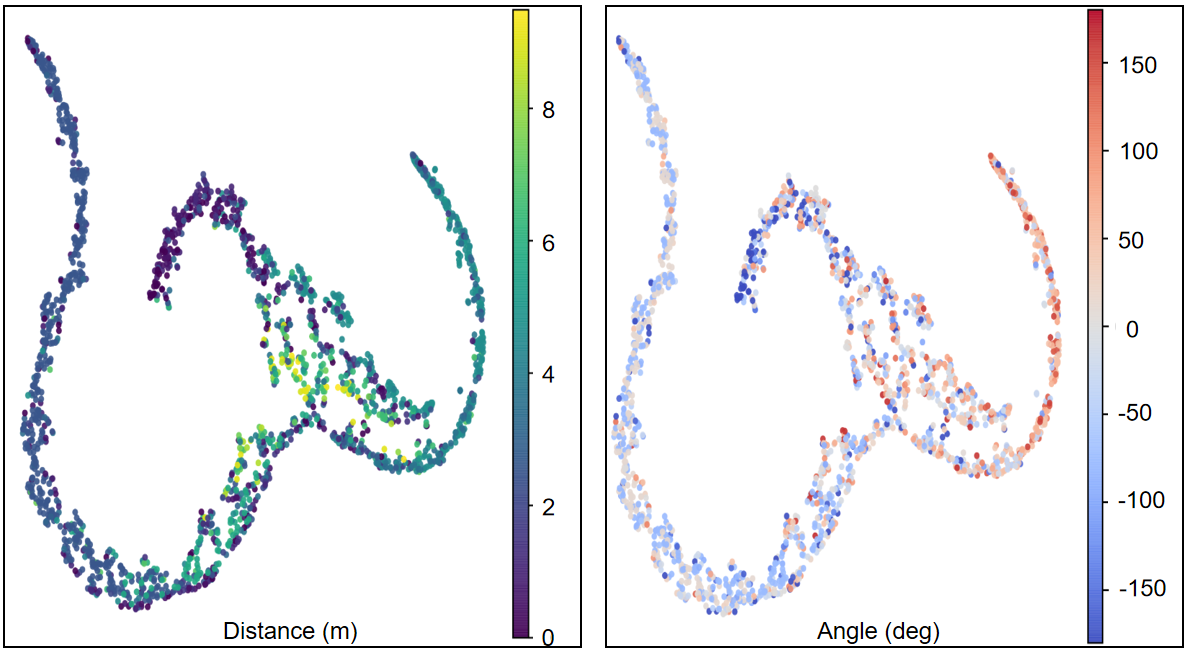}
    \caption{t-SNE projection of the audio features extracted by the ASE module. The left figure is colored according to the distance to the target, while the right figure is colored according to the relative angle to the target.}
    \label{fig:tsne}
\end{figure}

\subsection{Comparison with Baselines}\label{subsec:comparison}
The results of the comparative experiments are shown in Table~\ref{tab:main}. In the Heard task on the Replica~\cite{b7} and Matterport3D~\cite{b8} datasets, our model outperforms the SoundSpaces~\cite{b6} baseline on all metrics, indicating that the model has stable and effective navigation ability in the Heard task. In the more challenging Unheard task, our model shows significant advantages. Especially in the Replica~\cite{b7} scenarios, compared with the optimal method, the SPL, SR and SNA are improved by 28.6\%, 23.7\% and 9.8\% respectively. In the more complex Matterport3D~\cite{b8} dataset, the SR and SPL also achieve obvious improvements. The above results show that the proposed model can still achieve efficient navigation under unknown sound conditions. Its performance improvement does not come from the model's memory of specific sound directions and scenes, but from mining the spatial state information existing in the audio and realizing robust cross-modal fusion on this basis, thus reflecting good generalization ability under different environments and sound distributions.

\subsection{Ablation Study}\label{ablation}
In our study, we conducted ablation experiments on the Unheard task of both datasets, with the model evaluated under the following three settings: 
\begin{enumerate}
\item Removing both the ASE module and the ASGF module; 
\item Removing the ASE module and using a learnable vector as the query for cross-modal fusion; 
\item Removing the ASGF module, and using the spatial state information learned by ASE together with the visual and audio features extracted by the convolutional neural network as the decision input.
\end{enumerate}

The quantitative results are summarized in Table~\ref{tab:ablation_unheard}.

It can be observed that after removing the ASGF module, the SNA, SR and SPL all decrease significantly, which indicates that cross-modal spatial guidance and feature fusion play an important role in navigation decision-making under unknown sound conditions. Under the Unheard task, removing the ASE module makes it difficult for the agent to adapt to unknown sound source distributions, leading to a significant degradation in performance, which demonstrates that the effective utilization of spatial state information in audio is an important prerequisite for stable navigation in the Unheard scenario. Overall, the ablation experiments show that the ASE module extracts spatial state information from audio to provide key geometric constraints for navigation, while the ASGF module effectively fuses this information with visual observations, thus enabling the model to maintain stable and efficient navigation performance under unknown sound conditions. This verifies the good generalization ability of the proposed method under both unseen environments and unseen sound settings.

\subsection{Visualization Analysis}\label{FF}
To further analyze the representations learned by the ASE module, we performed t-SNE visualization on its output high-dimensional representations, as shown in Fig.~\ref{fig:tsne}. The figure displays 2000 samples extracted from the Unheard task on the Replica dataset~\cite{b7}. The left subfigure is colored according to the distance to the target. The right subfigure is colored according to the relative azimuth to the target. We can observe that the outputs of the ASE module do not show a random distribution in the embedding space. Instead, they form distinct continuous structures. When colored by the target distance, the samples exhibit a smooth color gradient along these structures. This suggests that the extracted features can continuously reflect the target distance information. When colored by the relative azimuth, the same continuous structures show consistent angular variation patterns. This demonstrates that the direction information is also effectively encoded. The above phenomena show that the ASE module learns target-related latent audio representations that preserve spatially informative cues, which can provide useful guidance for cross-modal fusion and navigation decision-making under unknown sound conditions.

We compared the top-down trajectories of ASGF-Nav and SoundSpaces in the 3D environment, as shown in Fig.~\ref{fig:trajectory}. From the trajectory plots, we can see that the navigation paths of ASGF-Nav are generally closer to the shortest path compared with the baseline. This result corresponds to its advantages in the SPL metric. In the more complex Matterport3D dataset~\cite{b8}, this difference becomes even more obvious. The trajectories of SoundSpaces~\cite{b6} often deviate from the optimal route. They also involve multiple direction corrections. In contrast, ASGF-Nav is more direct. It reduces unnecessary exploration processes. The trajectory comparison indicates that our method can achieve reasonable path planning even in complex environments. It also demonstrates better generalization ability.

\section{Conclusion}
We propose ASGF-Nav, a novel audio-visual navigation framework that leverages audio-derived spatial cues to guide dynamic multimodal fusion for improved generalization. Specifically, an Audio Spatial State Encoder is introduced to focus on reliable audio frames and learn informative latent audio representations, while an Audio Spatial State Guided Cross-Modal Fusion module adaptively balances auditory and visual contributions during decision-making. In addition, a gating mechanism is employed to suppress unreliable perceptual signals and enhance decision stability. Experimental results show that ASGF-Nav achieves superior navigation accuracy and path efficiency in complex unseen environments, demonstrating stronger adaptability and generalization. Overall, the results suggest that incorporating task-relevant audio representations into a reliable multimodal fusion mechanism is effective for audio-visual navigation.

\section*{ACKNOWLEDGMENT}

This research was financially supported by the National Natural Science Foundation of China (Grant No.: 62463029).

\bibliographystyle{IEEEtran}
\bibliography{reference}

\end{document}